\documentclass[epj]{svjour}
\usepackage{graphicx}
\begin{document}
\title{An experiment for the measurement of the bound-beta decay of the
free neutron} 

\author{W. Schott\inst{1}\and G. Dollinger\inst{2}\and T. Faestermann\inst{1}\and J. Friedrich\inst{1}%
\and F. J. Hartmann\inst{1}\and R. Hertenberger\inst{3}\and N. Kaiser\inst{1}\and A.~R.~M\"uller\inst{1}%
\and S. Paul\inst{1}\and A. Ulrich\inst{1} }

%
%
\institute{Physik- Department, Technische Universit\"at M\"unchen, D-85748 Garching,
Germany \and Institut f\"ur Angewandte Physik und Messtechnik,
Universit\"at der Bundeswehr M\"unchen, D-85577 Neubiberg, Germany
\and Sektion Physik der Ludwig-Maximilians-Universit\"at 
M\"unchen, D-85748 Garching, Germany}
\date{Received: date / Revised version: date}
%
\abstract{ The hyperfine-state population of hydrogen after the bound-beta decay
of the neutron directly yields the neutrino left-handedness or a
possible right-handed admixture and possible small scalar and tensor
contributions to the weak force. Using the through-going beam tube
of a high-flux reactor, a background free hydrogen rate of ca. 
3\,s$^{-1}$ can be obtained. The detection of the neutral hydrogen
atoms and the analysis of the hyperfine states is accomplished by
Lamb shift source type quenching and subsequent ionization. The
constraints on the neutrino helicity and the scalar and tensor
coupling constants of weak interaction can be improved by a factor of ten.
\PACS{13.30.Ce, 14.20.Dh} 
} 
\maketitle
\section{Introduction}
\label{intro} The neutron decay is for many years subject of intense
studies, as it reveals detailed information about the structure of
the weak interaction. However, the studies have so far only
addressed the main decay channel (classical neutron three-body
$\beta$-decay), where decay rates and decay asymmetries have been
determined with great precision \cite{Eid}. Symmetries of the
interaction are accessible by the precise measurement of
momentum spectra of the decay products and/or their correlation with
the neutron spin alignment. In addition, experiments are planned,
where the polarization of the final-state particles (electron or
proton) is measured. However, there is a very elegant method to
measure very precisely the relative spin alignments of the
daughter products and their correlation. Using the two-body neutron
$\beta$-decay into a hydrogen atom (H) and an electron antineutrino
($\bar{\nu}$)

\begin{equation}  \label{nHnu}
n \rightarrow H + \bar{\nu}
\end{equation}

\noindent one can investigate the hyperfine population of the
emerging hydrogen atom. The challenge lies in the very small
branching ratio BR = $4\cdot10^{-6}$ of the total neutron
$\beta$-decay rate \cite{Nem}, which is the result of the small phase
space for the efficient coalescence necessary for H-formation. We expect
only states with zero angular momentum in the hydrogen atom to be populated;
these are mainly the 1s and the metastable 2s states with 83.2\% and 10.4\%
probability, respectively. This is because states of angular
momentum $l>0$ have a negligible expectation value in the center of
the potential. The residual 6.4\% $ns$ states with $n>2$ decay
within nanoseconds, mainly into the 1s state.

Using standard V-A theory the possible spin configurations emerging
from this bound-$\beta$ decay and the resulting hyperfine states of the
emitted hydrogen atom are given in table 1. 
\begin{table}[bt]
\caption{Spin projections $i$ in the neutron bound-$\beta$
decay. As a convention, the H moves to the right, the
$\bar{\protect\nu}$ to the left. Fe and GT mean
Fermi and Gamov-Teller transition, respectively. $W_i$ are the
populations according to pure V-A interaction
(cf. eqs. \protect\ref{Wx_1}-\protect\ref{Wx_3} of ref. \cite{Nem},
$\protect\chi=\protect g_V/g_A$), $F$ the
total spin (with hyperfine interaction) and $m_F$ the $F$
projection, $\vert m_S m_I\rangle$ the Paschen- Back state, where
$m_S$ and $m_I$ denote the $e^-$ and $p$ spin quantum numbers (+
means +1/2, \textit{i. e.} spin points to the right in the magnetic
quantization field direction) }\label{tab1} \centering
\begin{tabular}{c|c|c|c|c|c|c}
configuration \textit{i}& $\bar{\nu}$ & $n$ & $p$ & $e^{-}$ & trans. & $W_{i}$(\%) \\
\hline
1 & $\leftarrow$ & $\leftarrow$ & $\leftarrow$ & $\rightarrow$ & Fe/GT & $%
44.14\pm.05$ \\
2 & $\leftarrow$ & $\leftarrow$ & $\rightarrow$ & $\leftarrow$ & GT & $%
55.24\pm.04$  \\
3 & $\leftarrow$ & $\rightarrow$ & $\rightarrow$ & $\rightarrow$ & Fe/GT & $%
.622\pm.011$  \\
4 & $\rightarrow$ & $\leftarrow$ & $\leftarrow$ & $\leftarrow$ &
Fe/GT & 0.\\
2' & $\rightarrow$ & $\rightarrow$ & $\rightarrow$ & $\leftarrow$ &
Fe/GT & 0.  \\
1' & $\rightarrow$ & $\rightarrow$ & $\leftarrow$ & $\rightarrow$ &
GT & 0. \\
\end{tabular}\\
\vspace*{0.5cm}
\begin{tabular}{c|c|c|c}
configuration \textit{i}& $F$ & $m_F $ & $\vert m_S m_I\rangle$ \\ \hline
1 & 0,1 & 0 & $\vert +-\rangle$ \\
2 & 0,1 & 0 & $\vert -+\rangle$ \\
3 &  1 & 1 & $\vert ++\rangle$ \\
4 & 1 & -1 & $\vert --\rangle$ \\
2' &  0,1 & 0 & $\vert -+\rangle$ \\
1' &  0,1 & 0 & $\vert +-\rangle$ \\
\end{tabular}%
\end{table}

According to \cite{Nem} the
population probabilities $W_i$ of the various configurations \textit{i} can
be deduced to be

\begin{equation}  \label{Wx_1}
W_1 = \frac{(\chi-1)^2}{2(\chi^2+3)},
\end{equation}

\begin{equation}  \label{Wx_2}
W_2 = \frac{2}{\chi^2+3},
\end{equation}

\begin{equation}  \label{Wx_3}
W_3 = \frac{(\chi+1)^2}{2(\chi^2+3)},
\end{equation}

\noindent depending only on one variable
$\chi=(1+g_S)/(\lambda-2g_T)$, with $\sum_{i=1}^{3}{W_i} = 1$.
$\lambda$ is the ratio
\begin{displaymath}
\lambda = g_A/g_V = -1.2695\pm.0029 \; 
\end{displaymath}
$g_A, g_V, g_S, g_T$ are the axial, vector, scalar and tensor
coupling constants, respectively. Thus, by means of $W_i$ only a
combination of $g_S$ and $g_T$ can be measured. $g_S$ is obtained
from $W_i$ only if $g_T$ is known from somewhere else and vice
versa.

The V-A contribution to the emission of H in one of its hyperfine
spin states with $F = 1$ and $m_F=1$ (config. 3 of table 1) is
suppressed by about two orders of magnitude (cf. eq. \ref{Wx_3}).
Therefore, by measuring the population of this configuration in
relation to the configurations 1 and 2 (both $m_F=0$), the existence
of {\em S, P, T} weak interaction variants can be verified. While the
states $F = 1, m_F = \pm1$ (configurations 3 and 4) are pure eigenstates of
the Hamiltonian describing the $p-e^-$ interaction in an external
magnetic field $B_1$, linear combinations of the configuration 1 and
2 states with coefficients depending on $B_1$ yield the other two
eigenstates. Therefore, the configurations 3 and 4 are separated in
any case from configurations 1 and 2, whereas a separation of
configurations 1 and 2 is only possible, if the decays occur within
a magnetic field $B_1$ (cf. below).

The sensitivity of $W_{i}$ with respect to $g_S$ and $g_T$ is shown
in table 2 \cite{Nem2}, where the $W_{i}$ are calculated assuming
small admixtures of $g_S$ and $g_T$.

There are three configurations, 4, 1' and 2', which correspond to a
scenario with right-handed neutrinos. However, only configuration 4
is unique and leads to a HFS state not populated by any other
transition. The negative helicity $\bar{\nu}$ configurations 1' and
2' would contribute to configurations 1 and 2, respectively.

\begin{table}[tbp]
\caption{$W_i$(\%) for various $g_S$ and $g_T$.}\label{tab2}
\centering
\begin{tabular}{c|c|c|c}
config. i & $g_S = 0$ & $g_S = 0.1$ & $g_S = 0$\\
& $g_T =0$ & $g_T =0$ & $ g_T=0.02$\\
\hline
1 & 44.14 & 46.44 & 43.40   \\
2 & 55.24 & 53.32 & 55.82  \\
3 & .622 & .238 & .780  \\
4 & 0. & 0. & 0.\\
\end{tabular}%
\end{table}

A possible small contribution of negative helicity to the $\bar{\nu%
}$ would manifest itself by a non-zero value of $W_4$ in table 1. 
The population of configuration 4, predicted by a left-right 
symmetric V+A model, is given by \cite{Byr}

\begin{equation}  \label{W4}
W_4 = \frac{(x + \lambda y)^2}{2(1 + 3\lambda^2 + x^2 + 3\lambda^2
y^2)},
\end{equation}

\noindent with $x = \eta - \zeta$ and $y = \eta + \zeta$, where
$\eta $ is the mass ratio squared of the two intermediate charged
vector bosons and $\zeta$ the boson mass eigenstate mixing angle,
$\eta$ and $\zeta$ being $\eta < 0.036$ \cite{Gap05} and $|\zeta |< 0.03$
(C.L. 90\%) \cite{Mus05}, respectively. In this model the antineutrino helicity 
$H_{\bar{\nu}}$ becomes

\begin{equation}  \label{Hnu}
H_{\bar{\nu}} = \frac{1 + 3\lambda^2 - x^2 - 3\lambda^2 y^2} {1 +
3\lambda^2 + x^2 + 3\lambda^2 y^2}.
\end{equation}

\noindent The neutron decay is then mediated by right-handed
currents, the details  of which are absorbed in $\eta$ and $\zeta$.
The best data for neutrino helicities come from $\tau$ and $\mu$ decay
spectra and the extraction of  the Michel parameters. Typical
accuracies are in the order of  15\% \cite{SLD}. Similar accuracies
can be achieved using the B-coefficient in  neutron decay.

A certain background to the configurations 1 - 4, 1', 2' comes from
$s$ states with larger principal quantum number $n$, also originally
populated. These $ns$ states with $n>2$ subsequently decay into the
1s and 2s states by spontaneous emission of photons, where the spin
quantum number $m_S$ of the $e^-$ is changed. If only the 2s state
is used for the spin analysis, the $4s$ and higher-state population
yield a $2s$ background slowly converging with $n$; as an example
$W(4s \rightarrow 2s) = 3.07\cdot10^{-4}$ and $W(5s \rightarrow 2s)
= 2.18\cdot10^{-4}$ (cf. Appendix).  From the sum $W(4s \rightarrow
2s) + W(5s \rightarrow 2s) = 5.25\cdot10^{-4}$ a fraction of 44.2\%
(that is $2.32\cdot10^{-4}$) would contribute to configuration~4 as
background, and 55.2\% of the sum ($2.90\cdot10^{-4}$) would show
up in configuration 3. The latter background constitutes already 47\% of the
resulting $W_3$ rate, which contributes $6.2\cdot10^{-4}$ to the $2s$
population.

In order to improve the present $g_S, g_T$ and $H_{\bar{\nu}}$
accuracies (cf. below), the background due to optical $m_S$ changing
transitions from $n$s states with $n>2$ into the $2s$-state analyzed
must be eliminated, \textit{e.g.} by ionizing these $ns$ H-atoms using a
laser prior to their decay.

\section{Experiment}
\label{sec:1}
\begin{figure*}[hbt]
\begin{center}
\includegraphics[width= 17cm]{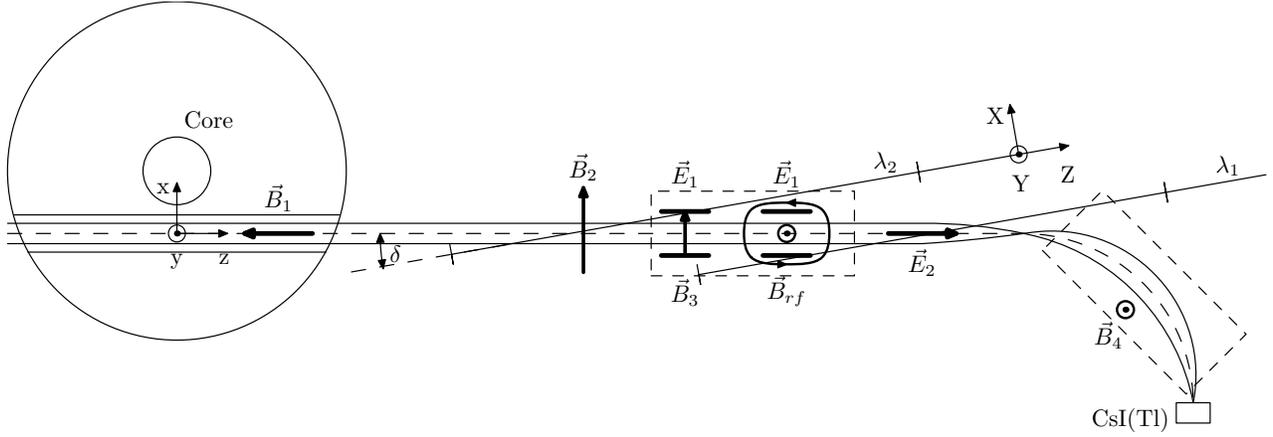}
\end{center}
\caption{Sketch of the experimental setup for measuring hydrogen 
atoms from bound-$\beta$ decay at FRMII. The through-going beam tube with the
magnetic quantization field $\vec{B_1}$ is drawn. Furthermore, a
field $\vec{B_2}$, rotating the
spin by $\pi$/2, a level splitting field
$\vec{B_3}$ with transverse electric field $\vec{E_1}$ and a
transverse rf magnetic field $\vec{B_{rf}}$ inside, a longitudinal
accelerating and focussing electric field  $\vec{E_2}$ and a bending
and focussing magnetic field $\vec{B_4}$ are indicated. By means of
the $\lambda_2 = 243$\,nm and the $\lambda_1 = 364$\,nm laser the
H-atoms are excited from $1s$ to $2s$ and ionized, respectively. They
are detected using a small CsI(Tl) crystal.} 
\label{FRMII}
\end{figure*}

Figure \ref{FRMII} depicts the suggested setup to perform
a neutron bound-$\beta$ decay experiment. Hydrogen atoms from this
decay have a kinetic energy of $T_{\mbox{H}}$ =
326.5\,eV, which corresponds to a velocity $v$ with $v/c =
0.83\cdot10^{-3}$. They are produced in the high neutron flux from
an intense neutron source and extracted from a through-going beam
pipe. In the following we will assume a high-flux neutron beam from
a reactor. Different possibilities for such a neutron source will be
discussed below.

The proposed setup of observing the emerging hydrogen atom parallel
to a magnetic field has the virtue that we automatically define a
helicity axis which coincides with the quantization axis for the
various HFS states. On the opposite side the neutrino helicity axis
follows from angular momentum conservation. With this, the spin
alignment of the original neutron is reconstructed. We thus do not
need to operate with polarized neutrons (see table 1). \ At the same
time we can reverse the direction of the quantization axis by
operating the system with reversed $B$ fields. This will become
important when discussing the spin analysis of the emerging hydrogen
by means of Paschen-Back splitting in a strong field.

The $B$ field in the decay volume is also necessary for a second
reason, which at the same time defines its minimum strength. In order
to preserve the magnetic quantum numbers of electron and proton,
$m_{S}$ and $m_{I}$, (also in the case of total spin $F=0$), the Zeeman
coupling to the external fields must be larger than the HFS coupling
within the hydrogen atom. The latter one corresponds to an internal
magnetic field of 507 Gauss (63.4 Gauss) for the $1s (2s)$ levels,
also called the critical field $B_{c}$ \cite{Hae}. Thus the field in
the decay volume should exceed the value of 63.4 Gauss, since the $2s$
level splitting is used.

During their passage in the beam tube the hydrogen atoms move to the right,
the hyperfine spin states $F=1,m_F=0,\pm1$ or $\vert m_S m_I\rangle$
states are kept in a longitudinal magnetic quantization field $B_1$,
which extends from the neutron-decay volume to the analyzing station. For
separating the configuration 1 and 2 states, $B_1$ should be at
least as large as the critical field $B_c$ of the $1s$ or $2s$
level($B_c$ is the magnetic field of the electron at the position of
the proton, which causes the hyperfine splitting (fig.
\ref{HF_split})). Also from this point of view, using only the $2s$
hydrogen atoms, $B_1$ must be $B_1 \ge 63.4\,$Gauss.

\subsection{Spin analysis and detection of 326.5\,eV H-atoms}
\label{sec:2}

\label{spin} The hyperfine analyzing part starts with a transverse
magnetic field $B_2$ = 10 Gauss. Within the $\vec{B_2}$ field region
of 5.4\,mm extension the $2s,F=1,m_F=0,\pm1$ states are rotated
adiabatically by $\pi/2$ into the same states, however, with the
quantization axis being
perpendicular to the direction of flight and parallel to the magnetic field $%
\vec{B_3}$ with $B_3$ = 575 Gauss, immediately following the
$\vec{B_2}$ field region.

\begin{figure}
\begin{center}
\includegraphics[width=8.0cm]{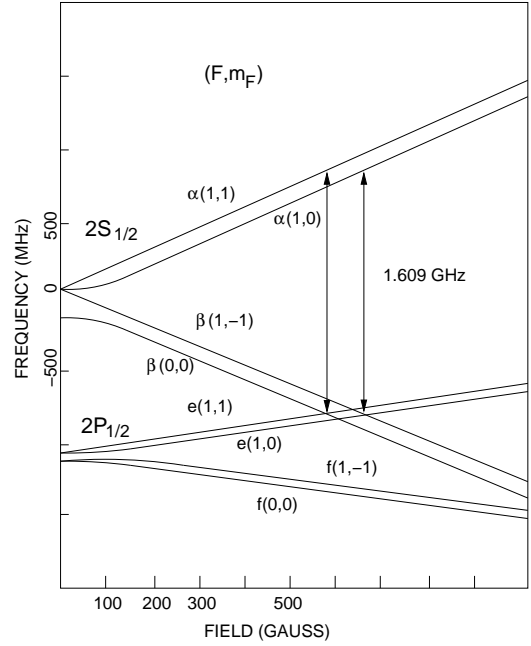}
\end{center}
\caption{Breit-Rabi diagram of the hyperfine splitting of the $2s$ and
$2p$ states in a magnetic field} 
\label{breit_rabi}
\end{figure}

Figure \ref{breit_rabi} depicts the Breit-Rabi diagram of the
$2s_{1/2}$ hyperfine states in a magnetic field, also showing the
corresponding $2p_{1/2} $ states \cite{Hae}.

The $\alpha$ and $\beta$ states are related to the $\vert m_S
m_I\rangle$ states as 

\begin{equation}
\vert \alpha11\rangle = \vert ++\rangle
\end{equation}
\begin{equation}
\vert \alpha10\rangle = \cos\theta\vert +-\rangle + \sin\theta\vert
-+\rangle
\end{equation}
\begin{equation}
\vert \beta1-1\rangle = \vert --\rangle
\end{equation}
\begin{equation}
\vert \beta00\rangle = \sin\theta\vert +-\rangle - \cos\theta\vert
-+\rangle,
\end{equation}

\noindent where $\theta$ is given by $\tan2\theta = B_c/B$ with the quantization field $%
B$ and the critical field $B_c$. 

If there exists a magnetic field $\vec{B_1}$ in the through-going
beam tube at the point of neutron decay, the $\vert \alpha10\rangle$
and $\vert \beta00\rangle$ population of the hydrogen atoms from 
bound-$\beta$ decay is given by

\begin{equation}  \label{Na10}
N_{\alpha10} = N_1 \cos^2\theta + N_2 \sin^2\theta
\end{equation}
and
\begin{equation}  \label{Nb00}
N_{\beta00} = N_1 \sin^2\theta + N_2 \cos^2\theta
\end{equation}

\noindent
with $N_1$ and $N_2$ being the configuration 1($\vert +-\rangle$) and 2($%
\vert -+\rangle$) state population from table \ref{tab1},
respectively. Thus, $\vert \alpha10\rangle$ and $\vert
\beta00\rangle$ are not equally populated if the bound-$\beta$ decays
occur in a magnetic field $B_1 \not= 0 $ (and, thus, $\theta \not=
\pi/4$). This population is conserved up to the region of spin
analysis (with $\vec{B_3}$ field) if no level transitions are
induced.

Owing to Stark mixing in an electric field superimposed transversely
to $B_3$, the states $\beta(1,-1)$ and $\beta(0,0)$, originally
metastable, can mix
with the $e(1,1)$ and $e(1,0)$ states at a $B_3$ field of 575 Gauss. Using 
$E_1 = 4.3$\,V/cm, the lifetime $\tau_{\beta}$ of the $\beta$ states
is thus strongly reduced ($v\cdot\tau_{\beta} = 1.1$\,cm) as compared
to the lifetime $\tau_{\alpha}$ of the $\alpha$ states(
$v\cdot\tau_{\alpha} = 1.8\cdot10^3$\,cm) \cite{Lam}. Thus, in the
$\vec{E_1}\times\vec{B_3}$ field the $2s_{1/2},F=1,m_F=-1$ and the
$2s_{1/2},F=0$ states are quenched immediately, decaying back into
the respective $1s_{1/2}$ ground state, whereas the two $2s_{1/2}$
states with $F=1,m_F=1$ and $F=1,m_F=0$ survive.

\begin{figure}
\begin{center}
\includegraphics[width=8.0cm]{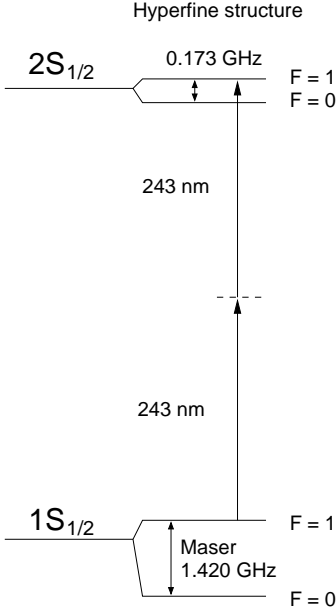} 
\end{center}
\caption{Hyperfine splitting of
the 1s and 2s state of hydrogen}
\label{HF_split}
\end{figure}

\begin{figure}
\begin{center}
\includegraphics[width=8.0cm]{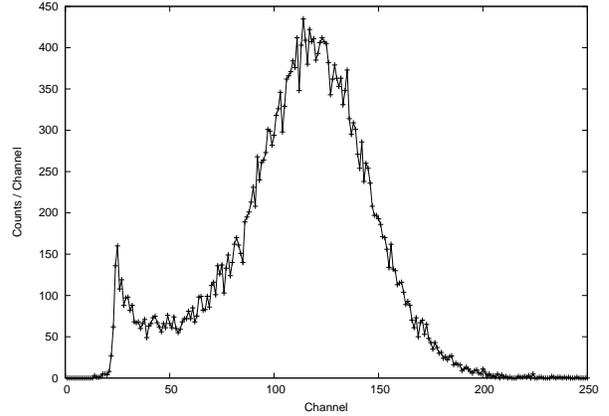}
\end{center}
\caption{Pulse-height spectrum from 18.5-keV protons measured with a 1\,cm$^3$ 
cubic CsI(Tl) crystal and a photomultiplier} 
\label{Charge}
\end{figure}

\begin{figure}
\begin{center}
\includegraphics[width=8.0cm]{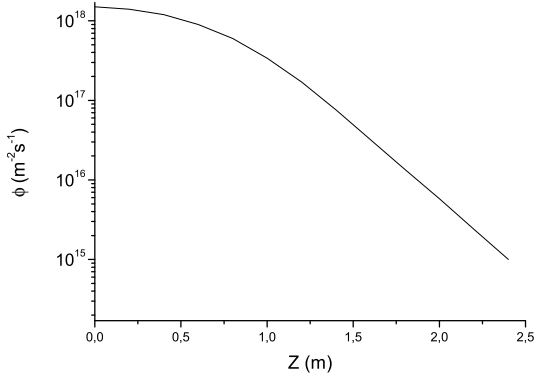}
\end{center}
\caption{Thermal neutron flux along the FRMII SR6 through-going beam
tube} 
\label{PHI_Z}
\end{figure}
One of the two remaining $\alpha$ states can be selected by the spin
filter method using the simultaneous interaction of the $\alpha$,
$\beta$ and $e$
levels at $B_3$ \cite{Hae}. Owing to the static transverse electric field 
$\vec{E_1}$ (fig. \ref{FRMII}), $\beta$ and $e$ are mixed. Then, we
apply an rf field of 1.609\,GHz, causing the $\alpha$ states to
interact with the coupled $\beta-e$ levels. At $B_3 \approx$ 
538\,Gauss the $\alpha(1,0)$ will be quenched and the $\alpha(1,1)$ state
will be selected, whereas with the same frequency at $B_3 \approx$
605\,Gauss the $\alpha(1,0)$ will remain, and the $\alpha(1,1)$ will be
removed. If the quantizing field $\vec{B_1}$ is reversed,
$\alpha(1,1)$ will be replaced by $\beta(1,-1)$ and $\alpha(1,0)$ by
$\beta(0,0)$, respectively. The transmission curve for a single spin
state is 0.4\,Gauss wide because of the width of the perturbed
$\beta$ state \cite{Hae,Eck}. Therefore, for a perfect
separation of the spin states, $B_3$ must be homogeneous to $\pm.2$
Gauss. The rf field can be produced within a box-shaped cavity in
the TE101 mode, yielding $\vec{B_{rf}}$ field lines perpendicular to
$\vec{B_3}$ (fig. \ref{FRMII}) surrounding the
cavity center $y$ axis corresponding to an rf electric field in the $\vec{B_3%
}$ direction and, thus, causing $\alpha-\beta$ transitions. Because of the $%
\beta-e$ mixing required at the position of $\vec{B_{rf}}$, the static $\vec{%
E_1}$ field must be implemented within the cavity.

As discussed in the introduction, the detection of a population of
configuration 4 from table 1, which feeds the state $\beta (1,-1)$,
is of eminent importance. The spin filter must thus be operated
such that only the $\vert --\rangle$
configuration survives, requiring a full quenching of the $\alpha $%
-states. This can be achieved by operating the experiment with
reversed
fields. The magnetic level splitting will thus be reversed in sign, where $%
\beta (1,-1)$ becomes $\alpha (1,1)$ and $\beta (0,0)$ becomes $\alpha
(1,0)$, respectively (however, we will keep the nomenclature defined
in the Rabi diagram of fig. \ref{breit_rabi}). A depopulation of the two
$\beta$-states requires a passage of the H-atoms through a varying
magnetic field with crossed $E$ field. When passing the fields of 538
and 605 Gauss, Stark mixing will cause a depopulation of the
corresponding 2s states. We are left with the two $\alpha $-states.
We then pass through the same filter as before operated at a field
of 538\,Gauss and a transition laser frequency of 1.609\,GHz. This
will cause a transition between $\alpha (1,1)$ and $\beta (0,0)$ with
subsequent quenching. We will be left with the desired configuration
$\vert --\rangle$ (table 1) which can be identified as before through
ionization and proton detection.

As the accuracy required is extremely large (zero measurement at a level of $%
10^{-6}$ - $10^{-7}$), we must take care of false effects. Some of
them are listed below.

\begin{itemize}
\item inefficient depopulation of the unwanted HFS states.
\item atomic cascades from $ns$ states (with $n>2$) which may feed $2s$ via an
intermediate $p$-state in an uncontrolled way. See the Appendix for
details.
\item radiative effects: additional soft photons may distort our arguments
on angular momentum conservation and lead to a population of $np$
states (with $n>2$). This can be calculated in the framework of
radiative corrections.
\end{itemize}

\noindent $\chi$ can be obtained by measuring the ratios\,
$v_{\alpha\beta} = N_{\alpha10}/N_{\beta00}$ or $v_{\alpha\alpha} =
N_{\alpha11}/N_{\alpha10}$ .

Downstream of the cavity and the $\vec{B_3}$ field the remaining
state-selected H-atoms (e. g., $2s_{1/2},F=1,m_F=1$) are ionized by, e. g., a 
$\lambda_1 = 364$\,nm laser beam (fig. \ref{FRMII}). Alternatively,
the ionization may be driven by an optical two-step process, using,
e. g., a $2s\rightarrow 3p$ transition ($\lambda =$ 656.28\,nm) and subsequent
ionization by a high-power laser ($\lambda =$ 816.33\,nm) or an
incoherent light source of that
wavelength. The produced protons are accelerated by an electrostatic field $%
\vec{E_2}$ to about 20\,keV energy and focussed to a small spot on
the beam axis, from where they are deflected by $90^o$ by an 
analyzing magnetic field and transported by point to point imaging onto a
CsI(Tl) scintillator. Using a photomultiplier to detect the light
output the protons can be detected with high efficiency as
demonstrated in a test experiment, where 18.5 keV protons
have been measured (fig. \ref{Charge}). A box-shaped magnetic field $\vec{%
B_4}$ is used for the bending, which provides radial and, owing to
non-zero edge angles, also axial focussing of the proton beam.

One interesting addition would be an intense $\lambda_2 = $243\,nm
laser beam crossing the beam line at an angle $\delta$ = 100 mrad
for exciting the H-atoms from $1s$ to $2s$ by means of two-$\lambda_2$
photon absorption (fig. \ref{HF_split}). Although not considered in
the present context, this pumping station would allow to use about
50\% of all hydrogen atoms produced in the subsequent hyperfine analysis,
as compared to the 10.4\% H-atoms from direct $2s$ population,
the basis for all following rate discussions. However, no high-power
Lyman-$\alpha$ laser is presently available.

In order to suppress the configuration 1-4 background due to $n$s
states with $n>2$, the corresponding H-atoms can be ionized by a
$\lambda = 816.33$\,nm laser with resonators positioned at both ends
of the experiments straight section, {\sl i.e.} at the left end of the 
through-going beam tube (fig. \ref{FRMII}) and behind the $\vec{E_2}$
focus.

\subsection{Possible event rates}

In the following we will evaluate possible event rates. We will
assume the set-up, as depicted in fig. \ref{FRMII}, to be installed at the
new Munich high-flux reactor FRMII. Extrapolations to other
neutron sources will be derived at the end.

The neutron decay volume within the beam pipe SR6 of the FRMII has a
length $l=2\cdot z_S$, with $z_S=4.7$\,m and a radius $r_S = 0.0715$\,
m. $z_S$ is the distance between center and far end of SR6. The
expected H-rate $\dot{N}_H$ at one exit of this
beam pipe, at a distance $z_S=4.7$\,m from its center (fig.
\ref{FRMII}), is given by

\begin{equation}  \label{H_rate}
\dot{N}_H = BR\cdot \int(\Phi(z)\Omega_S(z)\,dV)/4\pi\cdot 1/\tau_n
\cdot 1/v_n
\end{equation}

\noindent where $\tau_n$ = 886\,s is the neutron lifetime, $\Phi(z)$
the thermal neutron flux along the SR6 $z$ axis (fig.
\ref{PHI_Z}) \cite{Gau,Alt}, $V$ the SR6 volume, $\Omega_S =
A_S/(z-z_S)^2 = \pi r_S^2/(z-z_S)^2$ the solid angle of the hydrogen
spectrometer, with $A_S = 0.016\,\mbox{m}^2$ being the SR6 cross section,
$v_n = 2.2\cdot10^3\,\mbox{m/s}$ is the average thermal-neutron velocity.
Thus, the integral can be written as
\begin{eqnarray*}
\lefteqn{
(1/4\pi)\int\Phi(z)\Omega_S(z)\,dV=}&& \\
&=(1/4\pi)\displaystyle\int_{-z_S}^{z_S}\Phi(z)\cdot 
A_S^2/(z-z_S)^2 \,\cdot \,dz =&\\
&=  r_S^4\cdot\pi /4\,\cdot
\,\displaystyle\int_{-z_s}^{z_s}\Phi(z) /(z-z_S)^2\,\cdot\,dz. &(14)\\
\label{int_phi}
\end{eqnarray*}
\stepcounter{equation}
\noindent
$\dot{N}_H = 3\,\mbox{s}^{-1}$ 
is obtained using the thermal
neutron flux distribution of refs. \cite{Gau,Alt}(fig.
\ref{PHI_Z}). According to eqs. \ref{H_rate} and \ref{int_phi}, the
neutron density $N_n/V$, i. e. the number of neutrons $N_n$ in the
observed SR6 volume $V$, is given by
\begin{displaymath}
N_n/V=\int_{0}^{z_S}(\Phi(z)\,dz)/(v_n\, z_S)=2.4\cdot 10^8\,\mbox{cm}^{-3},
\end{displaymath}
being about $3\cdot 10^4$ times larger than the ultra-cold neutron
(UCN) density in the recently proposed UCN sources, e.g. ref. \cite{Tri}.

\subsection{Expected experimental constraints on $g_S$, $g_T$ and $H_{\bar{%
\protect\nu}}$}

\label{gS_constr} A small $g_S$ or $g_T$ contribution including
the sign may be measured via the population probability $W_3$. In \cite{Ade} a 
$1\sigma$ confidence level upper limit for the absolute value of $g_S$
is quoted to be $g_S \le 6\cdot10^{-2}$. In order to obtain for
$g_T=0$ an assumed value for $g_S = 6\cdot10^{-2}$ with the same
accuracy using bound-$\beta $ decay, the statistical error $(\delta
W_3)_{\mbox{stat}}$ must be $1\sigma$, where $\sigma$ is the standard
deviation. The difference
\begin{eqnarray*}
\Delta W_3=&&\\
&= |W_3(g_S=6\cdot10^{-2},g_T=0)-&\\
&-W_3(g_S=0,g_T=0)|=&\\
&= 2.54\cdot10^{-3}&(15)\\
\label{dW_3}
\end{eqnarray*}
\stepcounter{equation}
\noindent can be written as $\Delta W_3 = (\delta W_3)_{\mbox{stat}} +
2\left|(\delta W_3)_{\lambda}\right|$, where $(\delta W_3)_{\mbox{stat}}$
is the statistical error, and $(\delta W_3)_{\lambda}$ the
uncertainty due to the error on $\lambda$. Using $d\lambda =
2.9\cdot10^{-3}$ \cite{Eid}, we obtain $(\delta W_3)_{\lambda} =
(dW_3/d\lambda)d\lambda = -1.10\cdot10^{-4}$. Hence, for measuring
$W_3(g_S=6\cdot10^{-2},g_T=0) = 3.68\cdot10^{-3}$ with 68\,\% 
confidence, $(\delta W_3)_{\mbox{stat}} = 2.32\cdot10^{-3}$ corresponds to
$1\sigma$. $W_3$ can be written as $W_3 = N_3/N$, with $N =
\sum_{i=1}^{3}{N_i}$ the total number of counts and $N_i$ the 
individual number of counts for configuration $i$, respectively. 
$\sigma$ is then $%
\sigma \approx \sqrt{N_3}/N$ yielding $N = W_3/\sigma^2 =
684$. The measuring time required is $t = N/\dot{N}$, with
$\dot{N} = \dot{N_H}$. If all hydrogen atoms from bound-beta decay
moving along the beam tube were considered, $t$ becomes about 228\,s. As
only the $2s_{1/2}$ states can be used (see the discussion above) with
$\dot{N} \approx 0.1\, \dot{N_H}$, the necessary time to confirm the
present $g_S=6\cdot10^{-2}$ upper limit is 2280\,s.

The statistical error for $g_S$ can be written as

\begin{eqnarray*}
(\delta g_S)_{\mbox{stat}}&=& \displaystyle(\displaystyle\frac{\partial g_S} {\partial W_3})_{g_S=6%
\cdot10^{-2},\,g_T=0}\cdot (\delta W_3)_{\mbox{stat}}\\ &=& \displaystyle\frac{\lambda(\chi^2+3)^2}{%
-\chi^2+2\chi+3}\cdot \sqrt{\frac{W_3}{N}}.
\end{eqnarray*}

\noindent With $(\delta g_S)_{\mbox{stat}} = 6\cdot10^{-3} \; (\chi\approx
1/\lambda, W_3=3.683\cdot10^{-3}), N = 4.4\cdot10^{4}$ results,
which corresponds to 40\,h measuring time, when only the $2s_{1/2}$
state is used. This reduces the present $g_S$ upper limit by a
factor of ten.

Assuming a finite detection efficiency $\epsilon$, these times scale
with the factor $1/\epsilon$. Since $\epsilon \ge .1$ seems to be
realistic, the $g_S$ upper limit could be lowered by a factor of ten 
within a few weeks of measuring time.

According to eqs. \ref{W4} and \ref{Hnu} with $\zeta = 0$, i. e., $x
= y = \eta$, $W_4$ and $H_{\bar{\nu}}$ are approximately given by
\cite{Byr}
\begin{equation}
W_4 \approx \frac{\eta^2(1 + \lambda)^2}{2(1 + 3\lambda^2)}
\end{equation}

\begin{equation}
H_{\bar{\nu}} \approx 1 - 2\eta^2 = 1 - \frac{4(1 + 3\lambda^2)} {(1
+ \lambda)^2}\cdot W_4.
\end{equation}

\noindent The statistical error of $H_{\bar{\nu}}$ becomes

\begin{equation}
(\delta H_{\bar{\nu}})_{\mbox{stat}} = \frac{4(1 + 3\lambda^2)} {(1 + \lambda)^2}%
\cdot \sqrt{\frac{W_4}{N}}.
\end{equation}

\noindent 
Assuming $\eta$ = 0.036, we obtain $W_4 =
8.1\cdot10^{-6}$ and $H_{\bar{\nu}}~=~0.997$. This yields 
$N = 8.3\cdot10^{3}$
for
($\delta H_{\bar{\nu}})_{\mbox{stat}} = 1\cdot10^{-2}$, 
\textit{i.e.} 8\,h measuring time using only the
$2s_{1/2}$ state.

\subsection{Observables for $\protect\chi$}

$\chi$ is given by the ratios $v_{\alpha\beta}$ or
$v_{\alpha\alpha}$,
defined in section \ref{spin}, according to eqs. \ref{Wx_1} to \ref{Wx_3}
and \ref{Na10} to \ref{Nb00},

\begin{equation}
v_{\alpha\beta} = \frac{(\chi-1)^2\cos^2\theta +4\sin^2\theta} {%
(\chi-1)^2\sin^2\theta +4\cos^2\theta},
\end{equation}

\begin{equation}
v_{\alpha\alpha} = \frac{(\chi+1)^2} {(\chi-1)^2\cos^2\theta
+4\sin^2\theta},
\end{equation}

\noindent with $\chi$ either

\begin{equation}  \label{xvab}
\chi=1\pm 2\sqrt{\frac{\sin^2\theta - v_{\alpha\beta}\cos^2\theta} {%
v_{\alpha\beta}\sin^2\theta - \cos^2\theta}},
\end{equation}

\noindent or

\begin{equation}  \label{xvaa}
\chi=\frac{-(1+v_{\alpha\alpha}\cos^2\theta ) \pm
2\sqrt{v_{\alpha\alpha} (1-v_{\alpha\alpha}\sin^2\theta
\cos^2\theta)}} {1-v_{\alpha\alpha}\cos^2\theta }.
\end{equation}%

\noindent 
If $g_S$ and $g_T$ are small, the '-' sign holds in eq. \ref{xvab} 
and the '+' sign holdsin eq. \ref{xvaa}.

\subsection{Necessary power of the lasers}

The ionization of $2s$ hydrogen atoms can be obtained in a two-step process.
The excitation from the $2s$ to the $3p$ state by a $\lambda_1 = 656.28$
\,nm laser beam (fig. \ref{FRMII}) is selective to the moving H atoms
from bound-$\beta$ decay because of the Doppler shift. The
consecutive ionization can be achieved by an intense incoherent
816.33\,nm light source. The necessary power $\dot{Q_1}$ in the
$\lambda_1$ laser resonator is given by

\begin{equation}  \label{q1}
\dot{Q_1}=j_1\,E_1\,\delta X\,\delta Y.
\end{equation}

\noindent
$j_1$ is the photon current density, $E_1 = 1.889$\,eV the photon energy, 
$\delta X = 0.3$\,mm and $\delta Y = 0.1$\,m the $\lambda_1$ laser-beam 
waist width in $x$ and $y$ direction, respectively. The 
requested photon current $j_1$ can be written as

\begin{equation}  \label{j1}
j_1 = -\frac{\ln(1-P_1)}{\sigma_1\,\Delta t},
\end{equation}

\noindent where $P_1(2s\rightarrow 3p)$ is the desired $2s\rightarrow 3p$
excitation probability, $\sigma_1$ the $2s\rightarrow 3p$ photon-absorption
cross section and $\Delta t$ the exposure time, $\Delta t = \delta
X/(v\,\delta) = 12$\,ns. $v$ is the neutron-decay hydrogen-atom velocity
and $\delta =100$\,mrad the $\lambda_1$ laser resonator inclination
with respect to the beam axis $z$. $\sigma_1$ is given by

\begin{equation}  \label{sig1}
\sigma_1 = \frac{\lambda^3A_{ik}}{8\pi c}(\frac{\lambda}{d\lambda})
= 5.08\cdot 10^{-16}\,\mbox{m}^2,
\end{equation}

\noindent
with the Einstein transition coefficient \cite{Wie}

\begin{equation}  \label{Aik}
A_{ik} = \frac{6.67\cdot 10^{13}\cdot f_{ik}}{(\lambda(\mbox{nm}))^2},
\end{equation}

\noindent 
where the oscillator strength $f_{ik}$ is $f_{ik} = 0.641$
for the $2s\rightarrow 3p$ transition \cite{Omi}. The $\lambda_1$ laser resonator quality 
$\lambda_1/\mbox{d}\lambda_1$ is matched to the relative width
d$\lambda_1/\lambda_1$ which is necessary because of the Doppler
broadening d$\lambda_1$ caused by the thermal motion of the decaying
neutron, d$\lambda/\lambda = v_n/c = 0.73\cdot 10^{-5}$. The Doppler
shift $\Delta\lambda$ renders a high selectivity against thermally
moving H with $\Delta\lambda/\lambda = 0.83\cdot 10^{-3}$. A
laser power $\dot{Q_1} = 2.4\,W$ is needed for $P_1 = 0.8$.

In order to reduce the background from the population of higher $ns$ states 
in the neutron decay, a minimum power for the ionization laser with
$\lambda = 816.33$\,nm of $\dot{Q_1} = 216.7$\,W is required, as may
be derived using eqs. \ref{q1} to \ref{Aik}. Considering $W_4$,
the efficiency for the removal has to be $P_1 = 1-8.1\cdot 10^{-7}$,
thus closer to one by one order of magnitude than the value of $W_4 =
8.1\cdot10^{-6}$ to be measured (cf. subsection \ref{gS_constr}). We
assume d$\lambda/\lambda = 0.73\cdot 10^{-5}$, the oscillator
strength $f \approx 1$ for the transition $n \ge 3$ into
the continuum \cite{Omi}, the lifetime of the $3s$ state $\Delta t = \tau (3s) =
158.4$\,ns, the photon energy $E_1 = 1.512$\,eV and the waist widths
$\delta X = \delta Y = 0.1$\,m.

At present, we have only considered H atoms in the $2s$ state. A much
improved count rate could be obtained when also the $1s$ hydrogen
atoms could be used which make up 83\% of the resulting hydrogen
final states. However, this requires a pumping of all $1s$ to $2s$ states
of the hydrogen atoms before entering the spin selector. This can in
principle be achieved by means of a UV laser beam crossing the hydrogen 
path under a small angle $\delta$ (cf. fig. \ref{FRMII}). 
The $1s\rightarrow 2s$ transition requires a two-photon process.
We estimated $P(1s\rightarrow 2s)$ from the ionization probability
$P(1s \rightarrow \mbox{p})$, which has been calculated for the three-photon
process by \cite{Rot}.

The photon density for the two-photon absorption must be large,
requiring a small $\lambda_2$ laser beam waist $\delta X$ (fig.
\ref{FRMII}). $\delta X$ is limited by the uncertainty relation,
constraining the $\lambda_2$ laser beam angular width within the 
resonator to be $\delta\Theta\,\ge\,\lambda_2/(2\pi\delta X)$,
yielding $\delta\Theta\,\ge\,0.13\, \mbox{mrad}$ for $\delta X = 0.3$\,mm 
and the resonator length ($2\delta X/ \delta\Theta$) to be less 
than 4.6\,m. The passage time
$\Delta t$ of the H atoms through the $\lambda_2$ laser resonator
inclined by $\delta =100$\,mrad with respect to the beam $z$ axis
is 12\,ns.

The necessary power density $I_2$ for the $1s$ to $2s$ transition in the 
$\lambda_2$ laser resonator may be estimated from the calculated
ionization probability $P(1s\rightarrow p)=6.5\cdot10^{-4}$ for a
$\lambda_2$ laser
pulse with $\Delta t= 10$\,fs duration and $I_2=5\cdot10^{12}$\,W/cm$^2$
\cite{Rot}. Assuming $P(1s\rightarrow p) = (C_1\cdot I_2 \cdot \Delta
t)^3$, $C_1$, $C_1=1.73\cdot10^{-4}$s$^2$/kg results. 
$P_2=P(1s\rightarrow 2s)$ should be given as 
$P_2=(C_1\cdot I_2 \cdot\Delta t)^2$,
yielding $I_2=1.66\cdot10^{7}$\,W/cm$^2$ for $P_2=0.1$.
The high power is at the present technical limit for a
continuous beam covering an area of 
$\delta X\cdot\delta Y =0.3$\,mm$\cdot100$\,mm = 30\,mm$^2$.

\subsection{Background suppression}

The fast hydrogen atoms from the bound-$\beta$ decay of the neutron, emitted to the
right-hand side of fig.\ref{FRMII}, are observed through collimators outside the
through-going beam pipe. The collimators have to be placed such that
the wall of the inner tube is not visible by the detectors, because 
backscattered neutralized protons from normal $\beta$ decays would
produce a huge background. The setup also minimizes the neutron flux
at the end of the beam tube. The protons emitted to the left are
deflected by electric or magnetic fields such that they do not hit
the back end of the beam tube, where they could be backscattered
and neutralized producing a background flux of H atoms in the
analyzing direction. High-vacuum conditions ($\le 10^{-6}$\,hPa) are
also recommended to avoid neutralization of protons from three-body
neutron decay and to avoid scattering of the H atoms from bound-$\beta$
decay. The background, {\sl i.e.} the probability $P_3$ for neutron or
proton scattering at the rest-gas atoms in
the beam-tube vacuum should be less than 1\%. $P_3$ is given by 
$P_3=\sigma_3\,n_3\,\Delta z\,\le\,10^{-2}$, where
$\sigma_3=10^{-16}$\,cm$^2$ is a typical scattering cross section,
$n_3$ the atom density of the rest
gas and $\Delta z=10$\,m the vacuum chamber length, yielding 
$n_3\le 10^{11}$\,cm$^{-3}$: the beam line vacuum should be better than
$4.1\cdot 10^{-4}$\,Pa.

\section{Conclusion}

In this manuscript we outline an experiment to determine the scalar and
tensor contribution to the weak interaction including their signs as
well as the constraints on the left-handedness of the neutrino with
ten times higher precision than the present value. Currently, the upper
limit on $g_S$ is 
$\left\vert g_S\right\vert\le\,6\cdot10^{-2}$ (C.L.\,68\%) \cite{Ade},
whereas $|g_T/g_A|\,\le\,0.09$ (C.L.\,95\%) \cite{Boo}. 
From the measurements of
neutron $\beta$-decay asymmetry coefficients upper limits for the
left handed and right handed
boson mass squared and the absolute value of the boson mixing angle 
are determined to be $%
\eta \, <\, 0.036$ \cite{Gap05} and $\left|\zeta\right|<0.03$
(C.L. 90\%) \cite{Mus05},
respectively. In
the proposed experiment the upper limits of $g_S$ or $g_T$ and $\eta$ or $%
\zeta$ can be reduced by a factor of ten.

The issue is addressed by measuring the spin correlation between the
proton and the electron. They are defined with respect to their
direction of flight (measuring direction) by selecting different
hyperfine states from the bound-$\beta$ decay of the neutron.
Although the experimental requirements are challenging, we show a
pathway to perform such an experiment at the through-going beam tube
of a high-flux reactor like the FRMII. Slightly better
conditions do exist at the corresponding beam tube of the high-flux
reactor in Grenoble, as the usable neutron density is about four
times higher \cite{Kes}. Next generations of spallation neutron
sources like the SNS may be even better suited, if a through-going
beam tube is provided, since higher neutron densities are expected
and the pulsed neutron beam makes the detection of hydrogen atoms and the
background suppression easier.
%
%
%
%
\section{Appendix}
The most important contribution from higher excited states of
hydrogen to
the $2s$ hyperfine states are given by the contribution to
$W(4s\rightarrow 2s)$ changing $m_{S}$. This may be written as

\begin{eqnarray*}
W(4s\rightarrow 2s)&&\\
&=&2\cdot W(4s)\cdot W(4s\rightarrow 3p)\cdot W(3p\rightarrow
2s)\cdot\\
&&W(\Delta j=0)\cdot W(\Delta j=\pm 1) \\
&=&3.07\cdot 10^{-4}
\end{eqnarray*}
\noindent where $W(4s)=1.3$\% is the original $4s$ population.
\begin{eqnarray*}
W(4s\rightarrow 3p)=A_{4s3p}/(A_{4s3p}+A_{4s2p})=0.416 & {\rm and}& \\
W(3p\rightarrow 2s)=A_{3p2s}/(A_{3p2s}+A_{3p1s})=0.118&&
\end{eqnarray*}
\noindent is the $4s\rightarrow 3p$ and $3p\rightarrow 2s$
transition probability, respectively. $A_{4s3p}$ is the
corresponding $4s\rightarrow 3p$ Einstein transition coefficient
\cite{Wie}. The factor two arises from the interchangability of
the time-ordered transitions with $\Delta j=0$ and $\Delta j=\pm 1$.
Assuming the two transition probabilities for the $e^{-}$ spin
quantum number $m_{S}$ ($m_{S}$ changing with $\Delta j=0$ and
$m_{S}$ non-changing with $\Delta j=\pm 1$) to be equal, where
$j=l+s$ is the spin orbit coupling realized, the probability for
$\Delta j=0$ and $\Delta j=\pm 1$ photons of the $p\rightarrow s$
transition is given by the multiplicities of the Zeeman-split $p$ and
$s$ levels. They are $W(\Delta j=0)=2/5$ and $W(\Delta j=\pm 1)=3/5$.
\par

\noindent Correspondingly, the $m_{S}$ changing background
contribution from the $5s$ state is given by
\begin{eqnarray*}
W(5s\rightarrow 2s) &&\\
&=&2\cdot W(\Delta j=0)\cdot W(\Delta j=\pm 1)\cdot W(5s)\cdot \\
&&[W(5s\displaystyle\rightarrow 4p)\cdot W(4p\rightarrow
2s)\\
&&+ W(5s\rightarrow 3p)\cdot W(3p\rightarrow 2s)] \\
&=&2.18\cdot 10^{-4}
\end{eqnarray*}
\noindent with $W(5s)$ = 0.7 \%,
\begin{eqnarray*}
W(5s\rightarrow4p)&=&A_{5s4p}/(A_{5s4p}+A_{5s3p}+A_{5s2p}),\\
W(4p\rightarrow2s)&=&A_{4p2s}/(A_{4p2s}+A_{4p3s}+A_{4p1s}),\\
W(5s\rightarrow3p)&=&A_{5s3p}/(A_{5s3p}+A_{5s4p}+A_{5s2p}),\mbox{and}\\
W(3p\rightarrow2s)&=&A_{3p2s}/(A_{3p2s}+A_{3p1s}).
\end{eqnarray*}

\section{Acknowledgement}
We are especially grateful to Prof. J. Byrne for carefully reading
the manuscript and many valuable comments. We also want to thank 
Dr. A. R\"{o}hrmoser for calculating the neutron and gamma flux in the FRMII SR6
beam tube.

%
{}
\end{document}